# Changing views about remote working during the COVID-19 pandemic: Evidence using panel data from Japan

Eiji YAMAMURA[1]

Department of Economics, Seinan Gakuin University/ 6-2-92 Nishijin Sawaraku Fukuoka, 814-8511.

Email: yamaei@seinan-gu.ac.jp

Yoshiro TSUSTSUI

Department of Sociology, Kyoto Bunkyo University, Japan.

Email: tsutsui@econ.osaka-u.ac.jp

[1]Corresponding Author

Acknowledgements:

We would like to thank Editage (http://www.editage.com) for editing and reviewing this manuscript for language. This study was supported by Fostering Joint International Research B (Grant No.18KK0048) from the Japan Society for the Promotion of Science.

## Abstract

COVID-19 has led to school closures in Japan to cope with the pandemic. Under the state of emergency, in addition to school closure, after-school care has not been sufficiently supplied. We independently collected individual level data through internet surveys to construct short panel data from mid-March to mid-June 2020, which covered before and after the state of emergency. We analyse how the presence of school-aged children influences their parents' views about working from home. After controlling for various factors using a fixed effects model, we find that in cases where parents were workers, and the children are (1) in primary school, parents are willing to promote working from home. If children are (2) in junior high school, the parents' view is hardly affected. (3) Surprisingly, workers whose children are primary school pupils are most likely to support promotion of working from home after schools reopen.

Due to school closure and a lack of after-school care, parents need to work from home, and this experience motivated workers with small children to continue doing so to improve work-life balance even after schools reopen.

# 1. Introduction

How did COVID-19 change working styles in 2020?[1] According to previous studies, outbreaks of viral disease lead to school closures (e.g. Cauchemez et al. 2008; Cauchemez et al. 2014; Adda 2016). Similarly, schools in various countries have closed in response to COVID-19 (Baldwin & Mauro 2020). Consequently, the lifestyles of households with small children have changed. Childcare plays a critical role in a child's growth process when it stays at home[2]. The closure of schools has necessitated childcare at home, which increased parents' time for childcare[3]. In particular, workers having a child in primary school face difficulties because their child is less mature and needs care in the daytime. Availability of after-school childcare leads to the continuity of maternal labour supply in Japan (Takaku 2019) [4].

---

[1] In US counties, a lockdown to mitigate the spread of COVID-19 spread has led to an 8% increase in the number of people who stay at home (Brzezinski et al. 2020). In the UK, workers with lower-paying jobs are less able to work from home (Costa-Dias et al. 2020). COVID-19 caused economic stagnation, which has had greater repercussions on sectors with high female employment shares (Alon et al. 2020).

[2] Lack of after school childcare leads to an increased risk of skipping school and use of alcohol and drugs (Aizer 2004). Economic recessions cause teenagers' risky behaviors (Pabilonia 2017). A mother's absence reduces the time a child spends in school (Pörtner 2016).

[3] In the field of economics, many existing studies deal with parental time with children (e.g. Gutiérrez-Domènech 2010; Aguiar et. al. 2013; Gimenez-Nadal & Molina 2014; Morrill & Pabilonia 2015; Gorsuch 2016; Romanm & Cortina 2016; Bauer & Sonchak 2017).

[4] Existing studies deal with the relationship between childcare availability and maternal employment (e.g. van Gameren & Ooms

However, in addition to school closures, a state of emergency has been declared to mitigate the spread of COVID-19 in Japan. Hence, after-school childcare has not been sufficiently provided in the childcare market. Parents with small children are obliged to care for them by not going to their workplace. Inevitably, they seem to desire to work from home.

Primary and junior high schools were closed throughout Japan after March 2, 2020. Subsequently, a state of emergency was declared on April 7, and deregulated on May 25. In response to the deregulation, schools reopened in May. Japan differed from other countries that also implemented the policy of school closure because the Japanese government did not adopt the lockdown, which was one of the more stringent measures against COVID-19. In Japan, workers who choose to go their workplace are not penalised. Japanese workers can decide whether to go to their workplace or work from home to enable them to care for their child[5]. Sevilla and Smith (2020) collected individual-level data in the UK and found that COVID-19 changed the allocation of childcare compared with prior to COVID-19. They conducted a survey in May and asked workers with small children about work arrangements before and after the lockdown. One of advantages of the current study is that it conducted surveys five times with the same respondents. Hence, we can use the panel data to identify the same person's change of their view during the COVID-19 pandemic.

Workers do not have the freedom of being able to work from home due to work environments (Shimazu et al. 2020). Therefore, it seems plausible that workers did not work from home even though they would like to. Different from existing studies examining change of work style (Sevilla & Smith 2020; Yamamura & Tsutsui 2020a; Hatayama et al. 2020), we consider the subjective views about working from home by controlling work style. This study examines how workers with a child in primary school desired working from home as the situation changed in response to the spread of COVID-19. The main findings are as follows. First, parents with a child in primary school desired to promote working from home[6]. This tendency was observed especially after the deregulation of the state of emergency and reopening of schools.

The remainder of this paper is organised as follows. Section 2 presents an overview of the situation in Japan and outlines the design of the surveys. Section 3 describes the empirical method. Section 4 presents and interprets the estimated results. The final section provides some reflections and conclusions.

---

2009; Havnes & Mogstad 2011; Abe 2013; Asai et al. 2015; Brilli et al. 2016).
[5] In the US, UK, France, and Italy, which adopted the lockdown, neither a firm's manager nor employees themselves can decide whether to go to the workplace.
[6] Some studies analyse stay-at-home behaviour during the COVID-19 pandemic (Doganoglu & Ozdenoren 2020; Engle et al. 2020; Yamamura & Tsutsui 2020b).

## 2. Design of surveys and data

Figure 1 shows the changes in the daily number of people infected with COVID-19 during the period from March 1 to June 30. On February 27, the Japanese government requested schools to close beginning in March, although this was not obligatory. In response, various schools (primary and junior high) throughout Japan were closed from March 2. Therefore, parents with school-aged children were confronted with an unexpected situation in which their child did not go to school in the daytime. However, parents can outsource childcare for childcare services. In this situation, we initiated an internet survey and conducted the first wave on March 13.

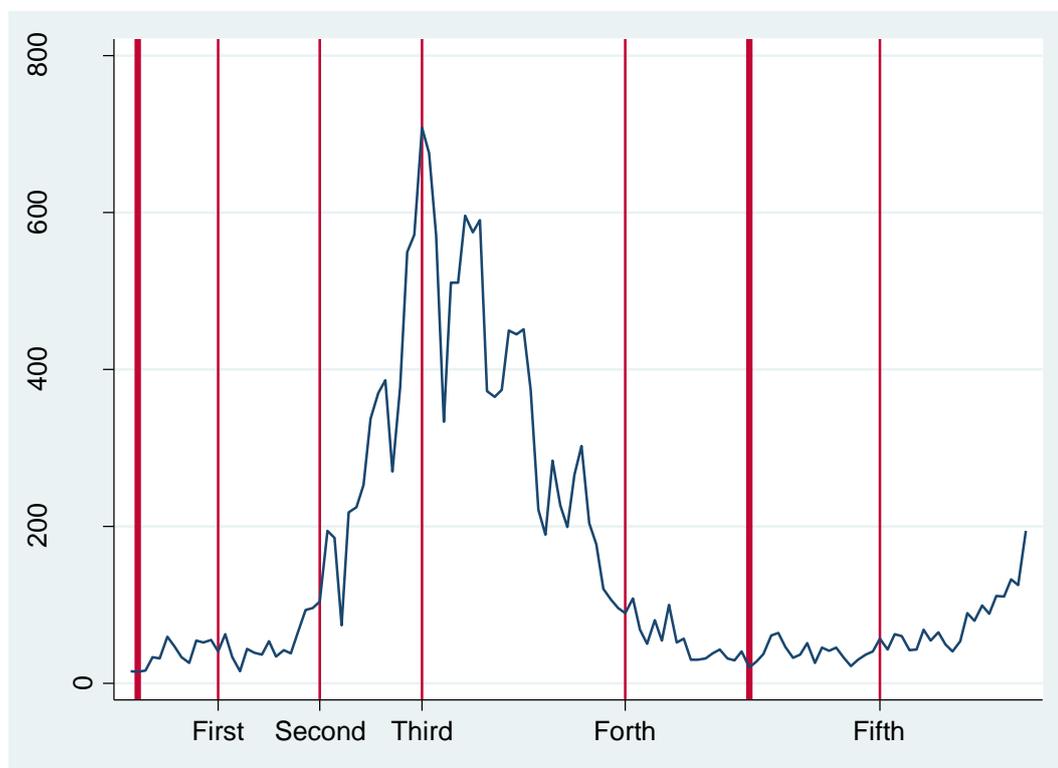

Figure 1. Changes in daily number of COVID-19 infections in Japan (March 1–July 2).

Note: First, second, third, fourth, and fifth waves were conducted on March 10, March 27, April 10, May 8, and June 12, respectively. Thin lines show the date of the surveys. Thick lines show the date when schools closures began (March 2) and when the state of emergency was deregulated (May 25). After the deregulation, schools reopened although the timing varied according to prefectures. A state of emergency was promulgated from April 7.

Source: Daily COVID-19 infections were sourced from the official website of the 'Ministry of Health, Labour and Welfare'. https://www.mhlw.go.jp/stf/covid-19/open-data.html. (On July 4, 2020).

As shown in Figure 1, the number of infected people drastically increased from the beginning of April. To cope with the rapid prevalence of COVID-19, the Japanese government declared a state of emergency on April 7. Similar to other countries (Baldwin & Mauro 2020), various public facilities such as museums and amusement parks were closed. However, the government only requested citizens stay home and avoid person-to-person contact and gathering together in enclosed spaces. These were not obligatory and there was no penalty if people did not follow the government request[7]. People could actually behave and make decisions in their daily lives based on their free will.

During the period of state of emergency, the pace of increase in the number of infections declined. On May 25, the state of emergency was deregulated. Consequently, schools reopened although there was variation of date of the reopening among regions. Schools were reopened in all parts of Japan on June 12 and the fifth wave of surveys was conducted on the same day.

## 2.1. Survey design

In February, even before COVID-19 spread in Japan, we planned to independently collect data to explore how COVID-19 influenced individual behaviours and households. We commissioned a research company to conduct surveys via the internet[8]. The sampling method was designed to collect a representative sample of the Japanese population about working style, views about working style, family members, job status, gender, age, educational background, and place of residence.

In the first wave of the survey, questionnaires were sent to selected Japanese citizens aged 16–79 throughout Japan and the same respondents participated in the subsequent waves. Figure 1 indicates that the surveys were conducted five time between March and June. Hence, we constructed short-term panel data.

The first wave was conducted on March 13. We gathered 4,359 observations, and the response rate was 54.7%. The second, third, fourth, and fifth waves were conducted on March 27, April 10, May 8, and June 12, respectively. The response rates reached 80.2% (second wave), 92.2% (third wave), 91.9% (fourth wave), and 89.4% (fifth wave). The sample was limited to workers because this study considered the preference for work style, this reduced the sample size to 8,903. Parents of primary school pupils were assumed to be younger than 50 years old, which was predicted from female child-bearing ages. Hence, we further limited the sample to respondents below 50 years old, which was also used for estimations. In addition to the sample of workers, we also used a sub-sample limited to workers under 50 years old.

---

[7] The situation was different from countries implementing drastic measures such as the 'lockdown' in the US, the UK, Italy, France, and Spain.
[8] INTAGE has extensive experience in academic research through internet surveys, and an excellent reputation.

## 2.2. Data

Table 1 presents descriptions of the variables used in this study. In waves 1–5, respondents were asked the following questions:

How do you consider *'the present condition of working from home'* as a countermeasure against COVID-19?

*Please answer in a scale from 1 (Sufficient) to 5 (Not sufficient).*

We define the answer as proxy variable for preference for working from home (*Work Home Preference*).

**Table 1**. Definitions of key variables and their basic statistics

|  | Definition | (1) With primary school pupil | (2) Others |
|---|---|---|---|
| *Work from home preference* | How do you consider *'the present condition of working from home'* as a countermeasure against COVID-19? Please indicate on a scale from 1 (Sufficient) to 5 (Not sufficient). | 3.74 | 3.64 |
| *Primary* | Equals 1 if respondent's child is in primary school pupil, 0 otherwise | 1 | 0 |
| *Junior High* | Equals 1 if respondent's child is in junior high school student, 0 otherwise | 0.17 | 0.08 |
| *Wave 1* | Equals 1 if survey is the first wave, 0 otherwise | 0.20 | 0.20 |
| *Wave 2* | Equals 1 if survey is the second wave, 0 otherwise | 0.20 | 0.20 |
| *Wave 3* | Equals 1 if survey is the third wave, 0 otherwise | 0.20 | 0.20 |
| *Wave 4* | Equals 1 if survey is the fourth wave, 0 otherwise | 0.20 | 0.20 |
| *Wave 5* | Equals 1 if survey is the fifth wave, 0 otherwise | 0.20 | 0.20 |
| *Schooling Years* | Respondent's years of schooling | 14.4 | 14.3 |
| *Income* | Respondent's annual household income. (Million yen) | 7.09 | 6.11 |
| *Ages* | Respondent's ages | 40.4 | 48.5 |
| *OLD Person* | Equals 1 if respondent has family member older than 80, 0 otherwise | 0.09 | 0.20 |
| *Female* | Equals 1 if respondent is female, 0 otherwise | 0.33 | 0.39 |
| *Infected COVID_19* | Number of persons infected by COVID-19 in the prefecture respondent resides. | 482 | 584 |
| *Remote work* | 'Within a week, to what degree have you achieved the not going to work? Please indicate on a scale from 1 (I have not achieved it at all) to 5 (I have completely achieved it)'. | 2.02 | 2.21 |

Note: The sample is limited to workers and excludes housewives, students, and retired persons.

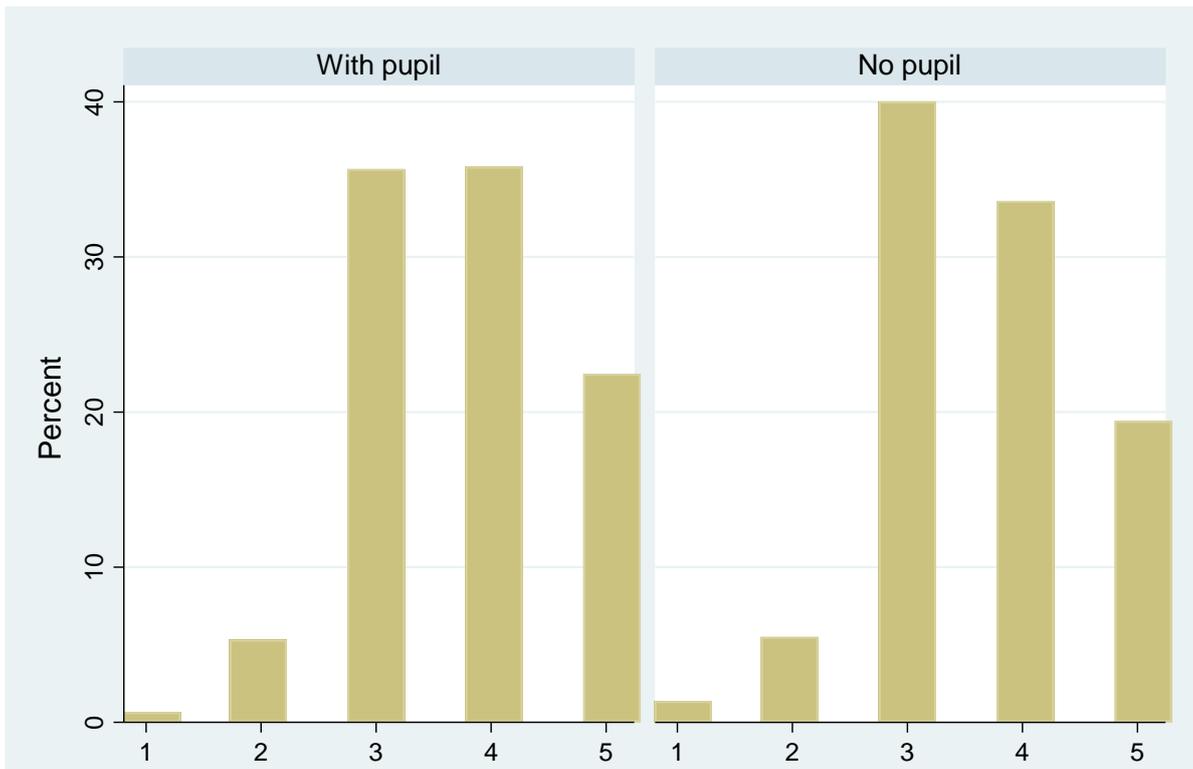

Figure 2. Distribution of *Work Home preference*

As shown in Table 1, *Work Home Preference* for workers with a child in primary school is greater than for other workers. This is consistent with observations in Figure 2 comparing the distribution of *Work Home Preference* between them. The findings indicate that parents of primary school pupils prefer working from home. We check how the effect of primary school pupil changes as the situation changes. Figure 3 presents mean values of *Work Home Preference* in each wave. We observe an increase in *Work Home Preference* from the first to the third wave, and a decrease after the third wave. Considering Figures 1 and 3 jointly indicates that respondents become more likely to prefer working until when the number of daily infections increased, but less likely to do so after flattening of the curve.

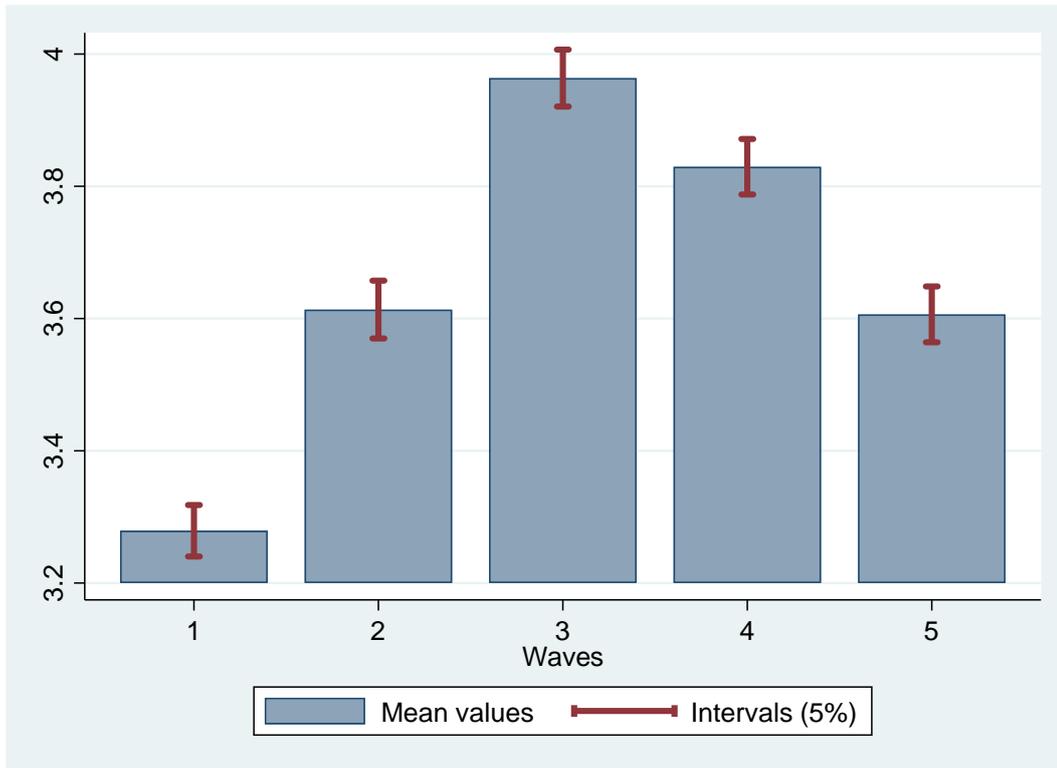

Figure 3. Changes in *Work Home Preference*

In waves 1–5, respondents were also asked the following question:

*'Within a week, to what degree have you achieved the not going to work?*

*Please indicate on a scale from 1 (I have not achieved it at all) to 5 (I have completely achieved it)'.*

The answer to this question is defined as the degree of working from home (*Remote Work*). Figure 4 shows its mean values in each wave. As shown, the degree of *Remote Work* increased from the first to the fourth wave, and decreased from the fourth to fifth wave. The degree of working from home did not reduce to the level before the state of emergency, even after it was deregulated. This implies that working from home is, to a certain extent, maintained regardless of the spread of COVID-19. According to Figure 3, workers are more likely to prefer working from home after the deregulation of the state of emergency than before its declaration. In our interpretation of comparing Figures 3 and 4, experience of working from home changes worker's preference for it. That is, through experience of working from home, workers become more likely to prefer working from home than before.

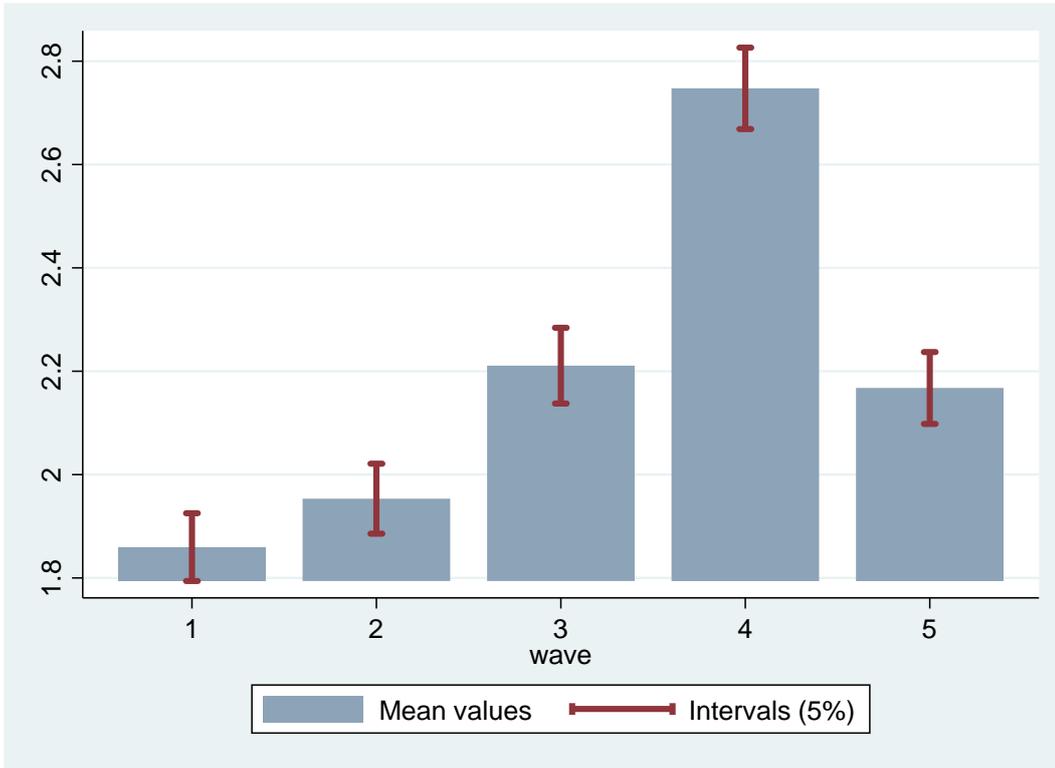

Figure 4. Changes in *Remote Work*.

## 3. Methodology

### 3.1. OLS model

We used a simple OLS regression model.[9] The estimated function takes the following form:

*Work Home Preference* $_{it}$ = $\alpha_0$ + $\alpha_1$ *Primary School* $_i$ + $\alpha_2$ *Junior High School* $_i$ + $\alpha_3$*Wave2* $_t$ + $\alpha_4$*Wave3* $_t$ + $\alpha_5$*Wave4* $_t$ + $\alpha_6$*Wave5* $_t$ + **X'B** + $u_{it}$, where *Work Home Preference* $_{it}$ represents the dependent variable for individual *i* and wave *t*. *Work Home Preference* is the key independent variable for exploring the effects of having children in primary school. To check for differences in the childcare requirements, *Junior High School*, which is a dummy for having a child in junior high school, is included in the estimations. The situation in Japan drastically changed during the study period, as illustrated in Figure 1. Therefore, we investigate impact of such change. Specifically, for testing the policy effect, we should pay attention to before and after the state of emergency. The second (*Wave 2*), third wave (*Wave 3*), fourth wave (*Wave 4*), and fifth wave (*Wave 5*) dummies are included; their reference group is the first wave. These dummies capture the degree of change in the dependent variables compared with the first wave. The regression parameters are denoted as *α*. **X** indicates the vector of control variables to capture demographic factors (*Ages, Female*), educational background (*Schooling*), economic condition (*Income*), family structure (*OLD Person*), and prevalence of COVID-19 *(Infected COVID_19)*. The error term is denoted by *u*. The data structure is a panel. However, we did not employ the fixed effects estimation because *Primary School* is constant and therefore captured as fixed effects. Accordingly, the estimation results of *Primary School* cannot be obtained.[10]

---

[9] *Work Home Preference* is an ordered, discrete variable. In this case, the ordered probit estimation is appropriate. However, its estimation results are similar to those derived from the OLS model. Our argument does not change when using the ordered probit model. However, the interpretation of the OLS results is simpler and easier to understand than that of the ordered probit model. Thus, we use OLS in this study. The results of the ordered probit model are available upon request from the corresponding author.

In addition to the sample of workers, we also use a sub-sample of workers under 50 years old because most parents of primary school pupils are thought to be under 50 years. We conduct the estimations using both the sample of workers and sub-sample of workers under 50 years old.

### 3.2. Fixed effects model

As alternative model for closer examination, we used the fixed effects model. The estimated function takes the following form:

*Work Home Preference* $_{it}$ =$b_1$ *Wave2* $_t$ ×*Primary* $_i$ + $b_2$ *Wave3* $_t$ ×*Primary* $_i$ + $b_3$ *Wave4* $_t$ ×*Primary* $_i$ + $b_4$ *Wave5* $_t$

×*Primary* $_i$ +$b_5$ *Wave2* $_t$ + $b_6$ *Wave3* $_t$ + $b_7$ *Wave4* $_t$ + $b_8$ *Wave5* $_t$ + $k_i$ + $u_{it}$,

In this specification, the fixed effects method was used to control for various time-invariant variables included in the baseline model such as *Schooling, Income, Age,* and *Old Person*. In addition, *Primary* $_i$ and *Junior High* $_i$ are also controlled although their cross terms were not controlled. $k_i$ captures the effects of various time-invariant variables. The cross term of wave dummies and *Primary* describes the degree of changing effect of having a child in primary school in the period. In alternative specifications, instead of *Primary*, the cross term of wave dummies and *Junior High* are included for checking the effect of having children in junior high school.

## 4. Results and interpretation

### 4.1. OLS estimation

Table 2 reports the results based on a sample of workers and sub-sample of workers under aged 50 in columns (1) – (2) and columns (3) – (4), respectively. In columns (2) and (4), *Remote Work* is included to control the degree of working from home because worker's preference depends on the actual degree of working from home. However, its results suffered from endogenous bias and we should therefore be cautious when interpreting the results of *Remote Work*.

---

[10] Under the Japanese educational system, it is possible that primary school students entered junior high school in April if they were in sixth grade in March. Similarly, junior high school students possibly entered junior high school in April if they were in third grade in March. However, we only asked respondents whether they have a child in primary school (junior high school) in Wave 1. Thus, we assume that *Primary School* and *Junior High School* are the same from Wave 1 to Wave 5.

**Table 2.** Baseline results (OLS model).

Dependent variables: *Work Home Preference*

|  | Full sample | | Ages<50 | |
|---|---|---|---|---|
|  | (1) | (2) | (3) | (4) |
| *Primary* | 0.09** | 0.09** | 0.07** | 0.07** |
|  | (0.03) | (0.03) | (0.03) | (0.03) |
| *Junior High* | −0.06 | −0.05 | −0.10 | −0.100 |
|  | (0.04) | (0.04) | (0.06) | (0.06) |
| *Wave 1* | < default > | | | |
| *Wave 2* | 0.33*** | 0.32*** | 0.33*** | 0.32*** |
|  | (0.02) | (0.02) | (0.03) | (0.03) |
| *Wave 3* | 0.67*** | 0.66*** | 0.65*** | 0.64*** |
|  | (0.02) | (0.02) | (0.02) | (0.02) |
| *Wave 4* | 0.49*** | 0.47*** | 0.45*** | 0.43*** |
|  | (0.04) | (0.04) | (0.04) | (0.05) |
| *Wave 5* | 0.26*** | 0.25*** | 0.25*** | 0.24*** |
|  | (0.05) | (0.05) | (0.05) | (0.05) |
| *Schooling* | 0.02*** | 0.02*** | 0.02** | 0.02** |
|  | (0.007) | (0.007) | (0.01) | (0.01) |
| *Age* | −0.002** | −0.003** | −0.002 | −0.002 |
|  | (0.001) | (0.001) | (0.002) | (0.002) |
| *Income* | 0.03 | 0.03 | 0.001 | 0.001 |
|  | (0.02) | (0.03) | (0.05) | (0.05) |
| *OLD Person* | −0.07 | −0.07 | −0.08 | −0.08 |
|  | (0.04) | (0.04) | (0.05) | (0.05) |
| *Female* | −0.006 | −0.009 | 0.04 | 0.03 |
|  | (0.029) | (0.03) | (0.04) | (0.04) |
| *COVID_19* | 0.05*** | 0.05*** | 0.06*** | 0.06*** |
|  | (0.006) | (0.007) | (0.008) | (0.009) |
| *Remote Work* |  | 0.03*** |  | 0.03** |
|  |  | (0.006) |  | (0.01) |
| R-Square | 0.08 | 0.08 | 0.08 | 0.08 |
| Obs. | 8,903 | 8,903 | 4,610 | 4,610 |

Note: Numbers within parentheses are robust standard errors clustered on residential prefectures. *** and ** indicate statistical significance at 1% and 5% levels, respectively.

As can be seen from Table 2, the coefficients of *Primary School* are positive and are statistically significant in all results. This suggests that in households with children in primary school, workers are more likely to prefer working from home than other workers. *Junior High School* does not show statistical significance in any column. This indicates that having children in primary school influences their parents' preference for working from home, whereas having children in junior high school does not. We interpret this as suggesting a difference in the necessity for childcare between children in primary school versus those in junior high school, who are considered to be more mature.

Regarding control variables, consistent with Figure 3, all wave dummies show a significant positive sign. Specifically, absolute values of their coefficients are the largest in third wave, which implies that working from home is most desired directly after the declaration of the state of emergency. *Schooling* was found to have a significant positive sign. This can be interpreted as higher educated workers are more able to work from home using internet technology, which leads them to prefer working from home. *COVID_19* produces a significant positive sign, which implies that the spread of COVID-19 leads people to desire to work from home. The significant positive sign of *Remote Work* shows a positive correlation between preference for working from home and the degree of working from home.

**4.2 Fixed effects estimation**

Table 3 reports the results of the fixed effects estimation. We observe the significant positive sign of *Wave3× Primary* and *Wave5× Primary*. This implies that workers with children in primary school are more likely to prefer working from home directly after the declaration of a state of emergency and after reopening of schools. Further, the value of its coefficients of *Wave5× Primary* is 0.16, which is larger than that of *Wave3× Primary* (0.12). This indicates that the gap of need to work from home between workers with children in primary school and other workers is larger by 0.16 in fifth wave and 0.12 in the third wave than that in the first wave. It is difficult to supply childcare services under the state of emergency. Therefore, workers want to work from home to be able to care for their child themselves. However, surprisingly, workers' with small children need for working from home is the highest after reopening school when they are less obliged to care for their children. As shown in Figure 4, the degree of working from home in the fifth wave is higher than that in second wave. Hence, workers are more likely to work from home after schools reopened than before the state of emergency. Meanwhile, Figure 3 shows that preference for working from home in the fifth wave is equivalent to that in second wave. Hence, workers' desire to work from home does not change between before the state of emergency and after reopening of schools. From the viewpoint of behavioural economics, people evaluate outcomes relative to a reference point and their preference therefore depends on this reference point (Kahneman & Tversky 1979). Our findings imply that experience of working from home changed workers' reference point.

Regarding control variables, we do not observe statistical significance for *COVID_19* and *Remote Work*. Different from Table 2, controlling for the fixed effects of respondents causes the statistical significance of *COVID_19* and *Remote Work* to disappear.

**Table 3.** Dependent variables: *Work Home Preference* (Fixed effects model).

|  | Full sample | | Ages<50 | |
|---|---|---|---|---|
| Wave 1 | < default > | | < default > | |
| Wave 2 × Primary | 0.07 (0.06) | 0.07 (0.06) | 0.08 (0.07) | 0.08 (0.07) |
| Wave 3 | 0.12* | 0.12* | 0.15* | 0.14* |

|  | (0.07) | (0.07) | (0.08) | (0.08) |
|---|---|---|---|---|
| × Primary | | | | |
| Wave 4 | 0.09 | 0.09 | 0.08 | 0.07 |
| × Primary | (0.07) | (0.07) | (0.08) | (0.08) |
| Wave 5 | 0.16** | 0.15** | 0.17** | 0.17** |
| × Primary | (0.07) | (0.07) | (0.08) | (0.08) |
| Wave 2 | 0.32*** | 0.32*** | 0.31*** | 0.31*** |
|  | (0.03) | (0.03) | (0.03) | (0.04) |
| Wave 3 | 0.67*** | 0.67*** | 0.63*** | 0.64*** |
|  | (0.03) | (0.03) | (0.04) | (0.04) |
| Wave 4 | 0.56*** | 0.57*** | 0.52*** | 0.53*** |
|  | (0.03) | (0.03) | (0.04) | (0.04) |
| Wave5 | 0.32*** | 0.32*** | 0.30*** | 0.31*** |
|  | (0.03) | (0.03) | (0.04) | (0.04) |
| Infected COVID_19 | −0.02* | −0.02* | −0.01 | −0.01 |
|  | (0.01) | (0.01) | (0.01) | (0.01) |
| Remote Work |  | −0.01 |  | −0.01 |
|  |  | (0.01) |  | (0.01) |
| Within R-Square | 0.13 | 0.13 | 0.12 | 0.12 |
| Groups | 1,655 | 1,655 | 946 | 946 |
| Obs. | 8,903 | 8,903 | 4,610 | 4,610 |

Note: Numbers within parentheses indicate robust standard errors clustered on individuals. ***, ***, * indicate statistical significance at the 1%, 5%, and 10% level, respectively.

In Table 4, we do not find statistical significance of cross terms between *Junior High* and wave dummies. Hence, workers with children in junior high school do not change their preference during the period.

**Table 4.** Dependent variables: *Work Home Preference* (Fixed effects model).

|  | *Full sample* | | *Ages<50* | |
|---|---|---|---|---|
| Wave 1 | < default > | | < default > | |
| Wave 2 | 0.01 | 0.006 | −0.03 | −0.03 |
| × Junior High | (0.08) | (0.08) | (0.10) | (0.09) |
| Wave 3 | 0.09 | 0.08 | 0.06 | 0.06 |
| × Junior High | (0.08) | (0.08) | (0.09) | (0.09) |
| Wave 4 | 0.05 | 0.05 | 0.02 | 0.01 |
| × Junior High | (0.08) | (0.08) | (0.09) | (0.09) |
| Wave 5 | −0.02 | −0.03 | −0.04 | −0.05 |
| × Junior High | (0.09) | (0.09) | (0.10) | (0.10) |
| Wave 2 | 0.33*** | 0.33*** | 0.33*** | 0.33*** |
|  | (0.03) | (0.03) | (0.03) | (0.03) |
| Wave 3 | 0.68*** | 0.69*** | 0.65*** | 0.67*** |
|  | (0.03) | (0.03) | (0.09) | (0.04) |
| Wave 4 | 0.57*** | 0.58*** | 0.54*** | 0.55*** |
|  | (0.03) | (0.03) | (0.04) | (0.04) |
| Wave 5 | 0.35*** | 0.35*** | 0.36*** | 0.36*** |
|  | (0.03) | (0.03) | (0.04) | (0.04) |
| Infected COVID_19 | −0.02* | −0.02* | −0.01 | −0.01 |
|  | (0.01) | (0.01) | (0.01) | (0.01) |
| Remote Work |  | −0.01 |  | −0.01 |
|  |  | (0.01) |  | (0.01) |
| Within R-Square | 0.13 | 0.13 | 0.12 | 0.12 |
| Groups | 1,655 | 1,655 | 946 | 946 |
| Obs. | 8,903 | 8,903 | 4,610 | 4,610 |

Note: Numbers within parentheses indicate robust standard errors clustered on individuals. ***, ***, * indicate statistical significance at the 1%, 5%, and 10% level, respectively.

Overall, the combined results in Tables 2–4 reveal that workers are likely to work from home if they have a child in primary school. Surprisingly, the gap in preference for working from home between workers with children in primary school and other workers is the largest after reopening of schools, rather than during the school closure period. These discoveries provide the compelling evidence that experience of working from home leads workers with small children to consider remote working and engaging in childcare to improve work-life balance.

## 5. Conclusion

Governments in various countries adopted the policy to close schools to cope with the COVID-19 pandemic in 2020. Under the state of emergency, in addition to school closures, childcare services have not been sufficiently supplied. Hence, parents are left with the burden of childcare. However, workers with small children have been confronted with difficulty in combining work and childcare during the day. We originally collected short-panel data covering before and after the state of emergency. Based on the data, we investigated whether the state of emergency leads workers with school-aged children to have the view to promote working from home. We found that parents with children in primary school to have the aforementioned view. However, the presence of children in junior high school did not influence it. Further, workers with children in primary school are most likely to support promotion of working from home than other workers after the deregulation of the state of emergency, and hence reopening schools.

This finding implies that closure of schools and after-school care during the day caused parents with children in primary school to require working from home. Moreover, these parents learn from the experience of working from home, which causes them to be aware of its effectiveness. Naturally, workers with small children are motivated to work from home to improve their work-life balance. They subsequently have view to support promoting working from home even after deregulation of the state of emergency and reopening of schools.

The findings of this study were based on Japanese data. Due to limitation of the data, it is unknown whether the argument holds in other countries that adopted more stringent measures against COVID-19 than Japan. It is therefore valuable for researchers to examine whether workers with small children are more likely to require working from home than other workers, which is an avenue for future research.

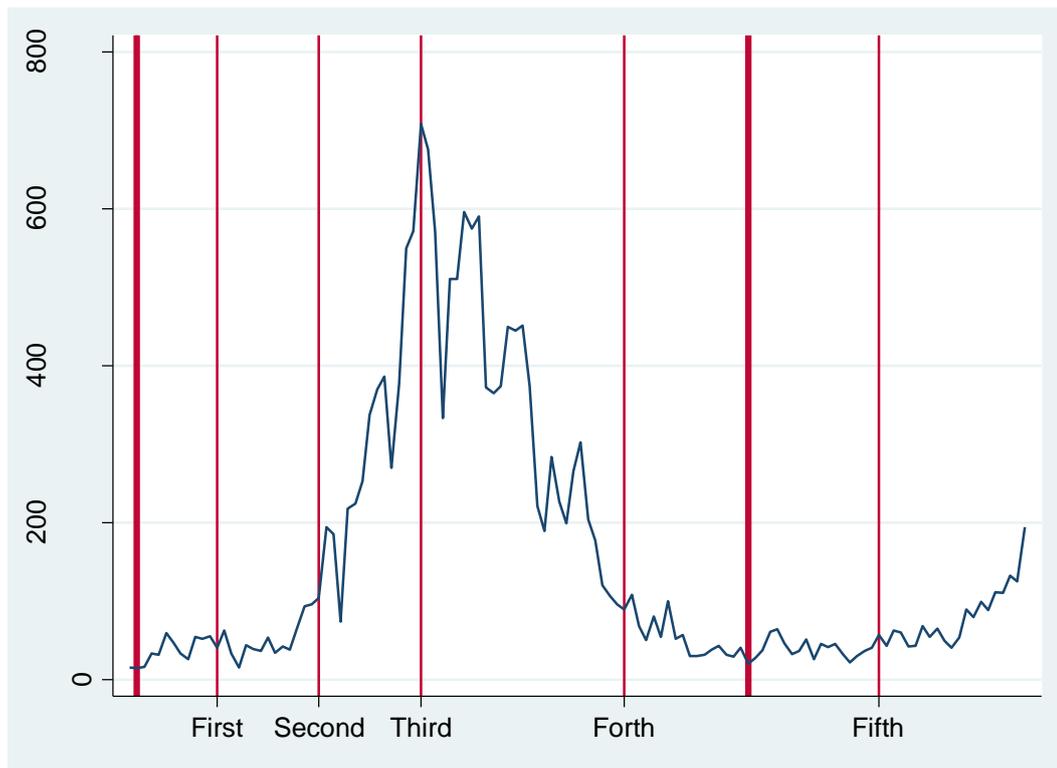

Fig 1. Changes in daily number of COVID-19 infections in Japan (March 1–July 2).

Note: First, second, third, fourth, and fifth waves were conducted on March 10, March 27, April 10, May 8, and June 12, respectively. Thin lines show the date of the surveys. Thick lines show the date when schools closures began (March 2) and when the state of emergency was deregulated (May 25). After the deregulation, schools reopened although the timing varied according to prefectures. A state of emergency was promulgated from April 7.

Source: Daily COVID-19 infections were sourced from the official website of the 'Ministry of Health, Labour and Welfare'. https://www.mhlw.go.jp/stf/covid-19/open-data.html. (On July 4, 2020).

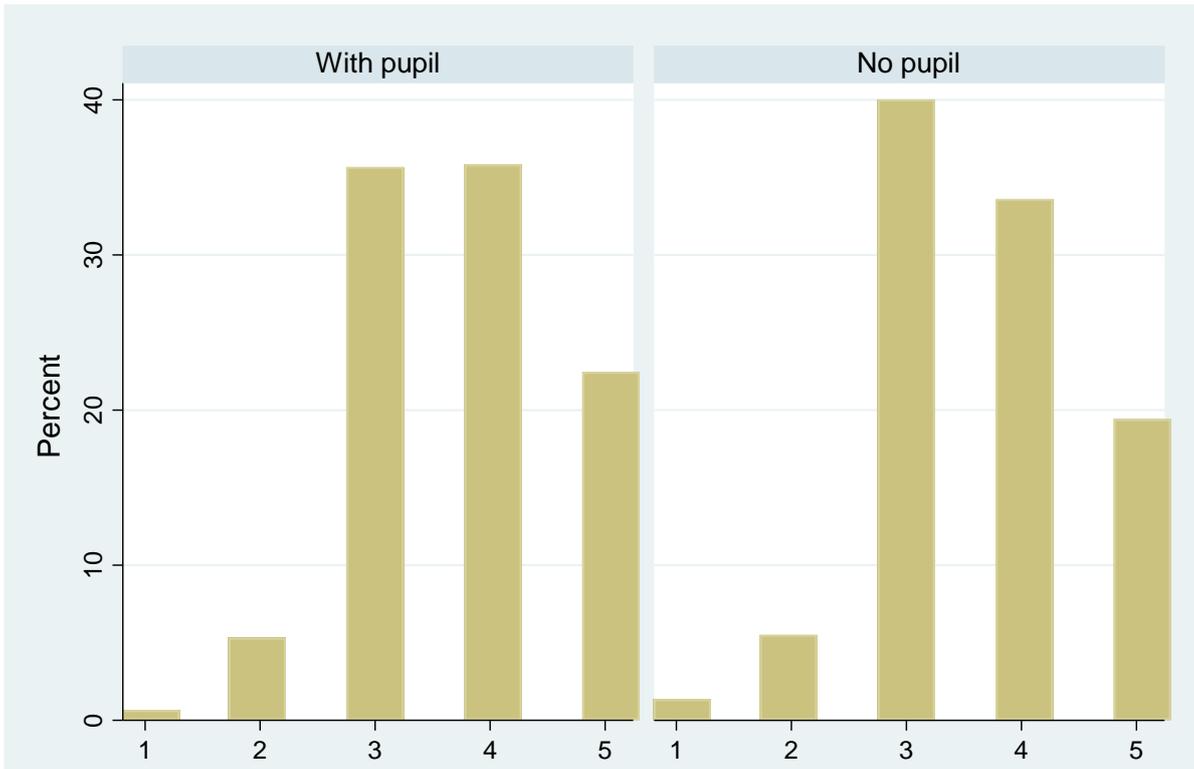

Figure 2. Distribution of *Work Home preference*

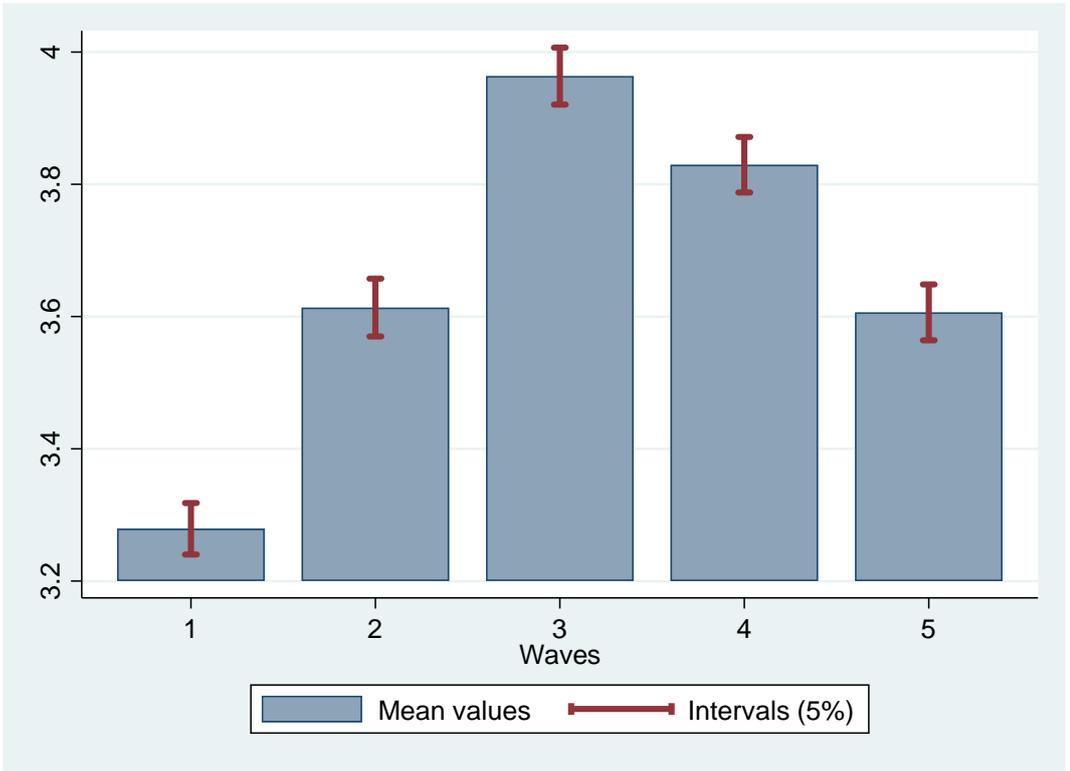

Figure 3. Changes in *Work home preference*

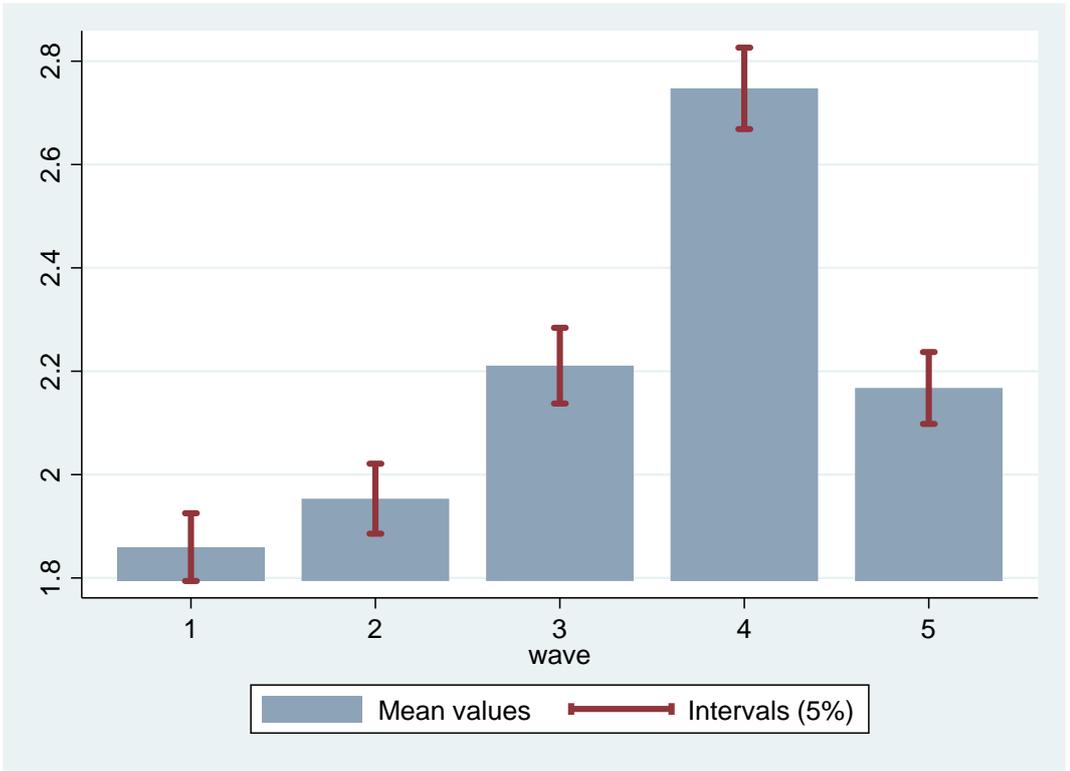

Figure 4. Changes in *Remote work*.

Table 1. Definitions of key variables and their basic statistics

| | Definition | (1) With primary school pupil | (2) Others |
|---|---|---|---|
| Work from home preference | How do you consider 'the present condition of working from home' as a countermeasure against COVID-19? Please indicate on a scale from 1 (Sufficient) to 5 (Not sufficient). | 3.74 | 3.64 |
| Primary | Equals 1 if respondent's child is in primary school pupil, 0 otherwise | 1 | 0 |
| Junior High | Equals 1 if respondent's child is in junior high school student, 0 otherwise | 0.17 | 0.08 |
| Wave 1 | Equals 1 if survey is the first wave, 0 otherwise | 0.20 | 0.20 |
| Wave 2 | Equals 1 if survey is the second wave, 0 otherwise | 0.20 | 0.20 |
| Wave 3 | Equals 1 if survey is the third wave, 0 otherwise | 0.20 | 0.20 |
| Wave 4 | Equals 1 if survey is the fourth wave, 0 otherwise | 0.20 | 0.20 |
| Wave 5 | Equals 1 if survey is the fifth wave, 0 otherwise | 0.20 | 0.20 |
| Schooling Years | Respondent's years of schooling | 14.4 | 14.3 |
| Income | Respondent's annual household income. (Million yen) | 7.09 | 6.11 |
| Ages | Respondent's ages | 40.4 | 48.5 |
| OLD Person | Equals 1 if respondent has family member older than 80, 0 otherwise | 0.09 | 0.20 |
| Female | Equals 1 if respondent is female, 0 otherwise | 0.33 | 0.39 |
| Infected COVID_19 | Number of persons infected by COVID-19 in the prefecture respondent resides. | 482 | 584 |
| Remote work | 'Within a week, to what degree have you achieved the not going to work? Please indicate on a scale from 1 (I have not achieved it at all) to 5 (I have completely achieved it)'. | 2.02 | 2.21 |

Note: Sample is limited to workers and excludes housewives, students, and retired persons.

Table 2. Baseline results (OLS model).

Dependent variables: *Work home preference*

|  | Full sample | | Ages<50 | |
| --- | --- | --- | --- | --- |
|  | (1) | (2) | (3) | (4) |
| *Primary* | 0.09** | 0.09** | 0.07** | 0.07** |
|  | (0.03) | (0.03) | (0.03) | (0.03) |
| *Junior High* | −0.06 | −0.05 | −0.10 | −0.100 |
|  | (0.04) | (0.04) | (0.06) | (0.06) |
| *Wave 1* | < default > | | | |
| *Wave 2* | 0.33*** | 0.32*** | 0.33*** | 0.32*** |
|  | (0.02) | (0.02) | (0.03) | (0.03) |
| *Wave 3* | 0.67*** | 0.66*** | 0.65*** | 0.64*** |
|  | (0.02) | (0.02) | (0.02) | (0.02) |
| *Wave 4* | 0.49*** | 0.47*** | 0.45*** | 0.43*** |
|  | (0.04) | (0.04) | (0.04) | (0.05) |
| *Wave 5* | 0.26*** | 0.25*** | 0.25*** | 0.24*** |
|  | (0.05) | (0.05) | (0.05) | (0.05) |
| *Schooling* | 0.02*** | 0.02*** | 0.02** | 0.02** |
|  | (0.007) | (0.007) | (0.01) | (0.01) |
| *Age* | −0.002** | −0.003** | −0.002 | −0.002 |
|  | (0.001) | (0.001) | (0.002) | (0.002) |
| *Income* | 0.03 | 0.03 | 0.001 | 0.001 |
|  | (0.02) | (0.03) | (0.05) | (0.05) |
| *OLD Person* | −0.07 | −0.07 | −0.08 | −0.08 |
|  | (0.04) | (0.04) | (0.05) | (0.05) |
| *Female* | −0.006 | −0.009 | 0.04 | 0.03 |
|  | (0.029) | (0.03) | (0.04) | (0.04) |
| *COVID_19* | 0.05*** | 0.05*** | 0.06*** | 0.06*** |
|  | (0.006) | (0.007) | (0.008) | (0.009) |
| *Remote Work* |  | 0.03*** |  | 0.03** |
|  |  | (0.006) |  | (0.01) |
| R-Square | 0.08 | 0.08 | 0.08 | 0.08 |
| Obs. | 8,903 | 8,903 | 4,610 | 4,610 |

Note: Numbers within parentheses are robust standard errors clustered on residential prefectures. *** and ** indicate statistical significance at 1% and 5% levels, respectively.

**Table 3.** Dependent variables: *Work home preference* (Fixed effects model).

|  | *Full sample* | | *Ages<50* | |
|---|---|---|---|---|
| *Wave 1* | < default > | | < default > | |
| *Wave 2* × *Primary* | 0.07 (0.06) | 0.07 (0.06) | 0.08 (0.07) | 0.08 (0.07) |
| *Wave 3* × *Primary* | 0.12* (0.07) | 0.12* (0.07) | 0.15* (0.08) | 0.14* (0.08) |
| *Wave 4* × *Primary* | 0.09 (0.07) | 0.09 (0.07) | 0.08 (0.08) | 0.07 (0.08) |
| *Wave 5* × *Primary* | 0.16** (0.07) | 0.15** (0.07) | 0.17** (0.08) | 0.17** (0.08) |
| *Wave 2* | 0.32*** (0.03) | 0.32*** (0.03) | 0.31*** (0.03) | 0.31*** (0.04) |
| *Wave 3* | 0.67*** (0.03) | 0.67*** (0.03) | 0.63*** (0.04) | 0.64*** (0.04) |
| *Wave 4* | 0.56*** (0.03) | 0.57*** (0.03) | 0.52*** (0.04) | 0.53*** (0.04) |
| *Wave5* | 0.32*** (0.03) | 0.32*** (0.03) | 0.30*** (0.04) | 0.31*** (0.04) |
| *Infected COVID_19* | −0.02* (0.01) | −0.02* (0.01) | −0.01 (0.01) | −0.01 (0.01) |
| *Remote work* |  | −0.01 (0.01) |  | −0.01 (0.01) |
| Within R-Square | 0.13 | 0.13 | 0.12 | 0.12 |
| Groups | 1,655 | 1,655 | 946 | 946 |
| Obs. | 8,903 | 8,903 | 4,610 | 4,610 |

Note: Numbers within parentheses indicate robust standard errors clustered on individuals.

***, ***, * indicate statistical significance at the 1%, 5%, and 10% level, respectively.

**Table 4.** Dependent variables: *Work home preference* (Fixed effects model).

|  | *Full sample* | | *Ages<50* | |
|---|---|---|---|---|
| *Wave 1* | < default > | | < default > | |
| *Wave 2* × *Junior High* | 0.01 (0.08) | 0.006 (0.08) | −0.03 (0.10) | −0.03 (0.09) |
| *Wave 3* × *Junior High* | 0.09 (0.08) | 0.08 (0.08) | 0.06 (0.09) | 0.06 (0.09) |
| *Wave 4* × *Junior High* | 0.05 (0.08) | 0.05 (0.08) | 0.02 (0.09) | 0.01 (0.09) |
| *Wave 5* × *Junior High* | −0.02 (0.09) | −0.03 (0.09) | −0.04 (0.10) | −0.05 (0.10) |
| *Wave 2* | 0.33*** (0.03) | 0.33*** (0.03) | 0.33*** (0.03) | 0.33*** (0.03) |
| *Wave 3* | 0.68*** (0.03) | 0.69*** (0.03) | 0.65*** (0.09) | 0.67*** (0.04) |
| *Wave 4* | 0.57*** (0.03) | 0.58*** (0.03) | 0.54*** (0.04) | 0.55*** (0.04) |
| *Wave 5* | 0.35*** (0.03) | 0.35*** (0.03) | 0.36*** (0.04) | 0.36*** (0.04) |
| *Infected COVID_19* | −0.02* (0.01) | −0.02* (0.01) | −0.01 (0.01) | −0.01 (0.01) |
| *Remote work* |  | −0.01 (0.01) |  | −0.01 (0.01) |
| Within R-Square | 0.13 | 0.13 | 0.12 | 0.12 |
| Groups | 1,655 | 1,655 | 946 | 946 |
| Obs. | 8,903 | 8,903 | 4,610 | 4,610 |

Note: Numbers within parentheses indicate robust standard errors clustered on individuals.

***, ***, * indicate statistical significance at the 1%, 5%, and 10% level, respectively.